# Edge states in self-complementary checkerboard photonic crystals: Zak phase, surface impedance and experimental verification

Xiao-Dong Chen, Ding Zhao, Xiao-Sheng Zhu, Fu-Long Shi, Huan Liu, Jin-Cheng Lu,

Min Chen, and Jian-Wen Dong*

*School of Physics & State Key Laboratory of Optoelectronic Materials and Technologies, Sun Yat-sen*

*University, Guangzhou 510275, China.*

*Corresponding author: dongjwen@mail.sysu.edu.cn

## ABSTRACT

Edge states of photonic crystals have attracted much attention for the potential applications such as high transmission waveguide bends, spin dependent splitters and one-way photonic circuits. Here, we theoretically discuss and experimentally observe the deterministic edge states in checkerboard photonic crystals. Due to the self-complementarity of checkerboard photonic crystals, a common band gap is structurally protected between two photonic crystals with different unit cells. Deterministic edge states are found inside the common band gap by exploiting the Zak phase analysis and surface impedance calculation. These edge states are also confirmed by a microwave experiment.

## I. INTRODUCTION

Photonic crystals (PCs) are composed of periodic optical structures in which the electromagnetic waves propagate in a similar way as the electrons move inside the periodic potential of semiconductor crystals [1-3]. By carefully designing PCs with various components and structures, the flow of light can be molded in anomalous ways and some fancy photonic phenomena can be observed, e.g., the super-prism [4], the negative refraction [5, 6], the sub-λ imaging [7], and the selective transmission [8]. Particularly, edge states in the photonic band gap of PCs have attracted much attention for the confinement and routing of light. For example, a photonic waveguide bend with near 100 percent transmission was achieved by joining three PC straight channels [9]. On-chip uni-directional propagation of spin-polarized light was realized by specially engineering the eigen-fields of edge states in glide-plane photonic crystal slabs [10]. Since the edge states are not only theoretically significant but also of application importance, it is desirable to establish the existing conditions of edge states based on firm theories. Recently, between two PCs with different topological invariants, e.g., Zak phase [11-14], Chern number [15-18], spin Chern number [19-22], valley Chern number [23-26], edge states are found due to the topological protection [27-29]. Geometric phase induced edge states were demonstrated in two mutually inverted PCs [30]. Besides, the deterministic edge states between two inverted semi-finite PCs with slightly disturbed conical dispersions at zone center were theoretically predicted [31] and experimentally observed [32].

As to have edge states between two PCs, one condition, i.e., a common band gap, should be satisfied. In this work, we consider the self-complementary checkerboard PC in which the "a common band gap" condition is naturally satisfied. Two topologically distinct PCs, i.e., PC1 and PC2, are constructed by different unit cells that are mutually complementary partners. To characterize the presence of edge states, different Zak phases of PC1 and PC2 are obtained and zero surface impedance for the photonic crystal boundary are found. A microwave experiment is also carried out to demonstrate these deterministic edge states.

## II. SELF-COMPLEMENTARY CHECKERBOARD PHOTONIC CRYSTAL

Let us start by considering the checkerboard PC shown schematically in Fig. 1(a). It is formed by two interlaced square lattices of dielectric rods with permittivity $\varepsilon_{diel}$ (black patches) and air rods (white patches). The lattice constant of PC is $a$, and the side length of dielectric/air rods is $b = a/\sqrt{2}$. As the checkerboard PC is invariant under the interchange of dielectric rods and air rods, it is self-dual or self-complementary [33]. Due to the self-complementarity, there are two ways to construct the checkerboard PC by choosing different unit cell. The first kind of unit cell is centered at the dielectric rod (outlined by the blue dashed square in Fig. 1(a)). The PC formed by periodically repeating the dielectric rod centered unit cell is shown in Fig. 1(b), and hereafter we name it as PC1. The other choice is that the unit cell is centered at the air rod (red dashed square). Similarly, PC2 is formed by periodically repeating the air rod centered unit cell [Fig. 1(c)]. Although PC1 and PC2 have different boundary morphologies, they have the same band structures as both their infinite structures form the checkerboard PC. As a result, the "a common band gap" condition will be naturally satisfied once a complete or directional band gap is found. Figure 1(d) shows the band structure for the transverse magnetic modes of the checkerboard PC with $\varepsilon_{diel} = 9$. Between the lowest two bands, there is a complete band gap ranging from $0.245c/a$ to $0.254c/a$ (shaded in green). This band gap is commonly shared by PC1 and PC2, and it serves as a good starting point to find edge states. Note that although the band width of the complete band gap is small (3.6%), the directional band gap is large enough for the good confinement of edge states. For example, the directional band gap along the ΓX direction has a 27.2% gap-midgap ratio.

# III. DETERMINISTIC EDGE STATES

## A. Zak phase analysis

To study the deterministic edge states at the photonic boundary between PC1 and PC2 (which are inherent from the checkerboard PC), we begin with the discussion about Zak phase. The reduced 1D band structures for PC1 and PC2 with $k_y = 0.25\pi/a$ are shown in Figs. 2(a) and 2(b), respectively. These band structures are exactly the same due to the self-complementarity of checkerboard PC. But their Zak phases are different. Here the Zak phase of the lowest band can be obtained by considering the symmetries of eigen-states at two high symmetry $k$-points in the reduced 1D Brillouin zone, i.e., $k_x = -\pi/a$ (labelled by A) and $k_x = 0$ (labelled by B) [30, 34]. Kohn's results [34] tell us that if either $|\tilde{E}_A(x=0)|$ or $|\tilde{E}_B(x=0)|$ is equal to zero while the other one is nonzero, the Zak phase is $\pi$; otherwise the Zak phase is 0. Here, $\tilde{E}_{\vec{k}}(x=0)$ is defined as $\tilde{E}_{\vec{k}}(x=0) = \frac{1}{a}\int_0^a E_{\vec{k}}(0,y)e^{-ik_y \cdot y}dy$ where the subscript $\vec{k}$ is considered at A or B points, and the eigen-fields at $x = 0$ are integrated. For example, the electric fields of two eigen-states for PC1 at A and B points are shown in the middle insets of Fig. 2(a). After the numerical summation, we found that $|\tilde{E}_A(x=0)| = 0$ while $|\tilde{E}_B(x=0)| \neq 0$. Therefore, the Zak phase is $\pi$, which is given along with the band structure shown in Fig. 2(a). On the other hand, the Zak phase is 0 for the band structure of PC2 shown in Fig. 2(b) as both $|\tilde{E}_A(x=0)|$ and $|\tilde{E}_B(x=0)|$ are nonzero. Hence, we get to the consequence that the first bands of PC1 and PC2 have different Zak phases. When the first lowest band gap above the first band is considered, deterministic edge states can be found at the boundary between such two PCs with different Zak phases [30]. To see this, we consider the photonic boundary shown in the inset of Fig. 2(c). On the left-hand side of this boundary is the semi-infinite PC1 while on the right-hand side is the semi-infinite PC2. This photonic boundary is periodic along the $y$ direction and its corresponding band structure is given in Fig. 2(c). Ingrained from the self-complementarity of checkerboard PC, a common band gap is found (shaded in white). Accompany with the appearance of this common band gap, edge states expanding from $k_y = 0$ to $k_y = \pi/a$ also appear (marked in green). Figure 2(d) plots the amplitude of electric fields of one representative edge state with $f = 0.223c/a$ at $k_y = 0.25\pi/a$ (marked by a green star in Fig. 2(c)). The edge state localizes around the boundary and decays exponentially away the boundary. For other values of $k_y$, edge states are also found in this photonic boundary with two constituent PCs having different Zak phases.

## B. Surface impedance calculation

Besides analyzing the Zak phase, the surface impedance calculation is another method to determine whether edge state will appear or not at the photonic boundary [35]. To see this, we obtain the surface impedance by the well-established retrieval method from scattering parameters [36, 37]. As schematically shown in Figs. 3(a) and 3(c), plane waves are incident with the wave-vector (**k**), and periodic boundary conditions are applied along the $y$ direction. In principle, a semi-infinite stack of layers should be employed along the $x$ direction, but in the simulation, the value of surface impedance inside the band gap converges to a constant when the number of layer is as many as 30. The reflection ($r$) and transmission ($t$) coefficients are exacted and the surface impedance Z can be obtained [36-38]:

$$Z(\omega, k_y) = \pm \frac{\sqrt{(1+r)^2 - t^2}}{\sqrt{(1-r)^2 - t^2} \cdot \sqrt{1 - k_y^2 / k^2}}, \quad (1)$$

where $k$ is the amplitude of the incident wave-vector, and $k_y$ is the tangential component along the boundary. The sign of surface impedance in Eq. (1) can be determined by the causality consideration: the real part of surface impedance, i.e., Re(Z) $\geq$ 0 should be satisfied. Inside the band gap, Re(Z) goes to zero and the imaginary part of surface impedance, i.e., Im(Z) determines how the waves reflect and transmit across the boundary. For example, Figure 3(b) plots the variation of Im(Z) for a given $k_y = 0.25\pi/a$ as a function of frequency. For PC1, the value of Im($Z_1$) decreases monotonically from 0 to about -4 with the increasing frequency inside the directional band gap ranging from $0.194c/a$ to $0.262c/a$ [see the blue curve in Fig. 3(b)]. Whereas for PC2, the value of Im($Z_2$) decreases monotonically from about 4 to 0 with the increasing frequency [see the red curve in Fig. 3(b)]. Due to the opposite signs between Im($Z_1$) and Im($Z_2$) and also their monotonicity, the zero surface impedance condition: Im($Z_1$) + Im($Z_2$) = 0 can be achieved at a particular frequency. At this frequency, one edge state of photonic boundary can be found. As marked by the green star at which Im($Z_1$) + Im($Z_2$) = 0 is fulfilled [Fig. 3(b)], the edge state exists at $f = 0.223c/a$, which is in good agreement with the frequency obtained by the full-wave calculations shown in Fig. 2(d).

For generality, Figures 3(d) and 3(e) respectively show Im($Z_1$) and Im($Z_2$) in the $\omega$-$k_y$ diagram. Here the surface impedance information for electromagnetic waves inside the directional band gap is given as we try to find edge states in the band gap. Also only the surface impedance information for the extended states locating above the light line is given as the evanescent waves below light line (shaded in grey) are not considered. For PC1, Im($Z_1$) decreases monotonically from 0 to -∞ with the increasing frequency, while Im($Z_2$) decreases monotonically from +∞ to 0 for PC2. As to find out the frequencies at which the zero surface impedance condition will be fulfilled, we plot Im($Z_1$) + Im($Z_2$) in Fig. 3(f). For each $k_y$, Im($Z_1$) + Im($Z_2$) drops monotonically from +∞ to –∞ with the increasing frequency. It will definitely go across the value of 0, and it implies that there exists one edge state inside the common band gap. This is confirmed by the white dashed curve in Fig. 3(f) where Im($Z_1$)

+ Im($Z_2$) = 0 is plot. It is in good agreement with the edge dispersion that is obtained by the full-wave calculations shown in Fig. 2(c).

### C. Microwave experimental verification

To prove the existence of edge states predicted by both the Zak phase analysis and surface impedance calculation, a microwave experiment is carried out. Figure 4(a) shows the photo of the experimental sample with PC1 locating on the left and PC2 on the right. The lattice constant is $a$ = 10 mm for both two PCs. The relative permittivity of dielectric squared rod (alumina) is $\varepsilon_{diel}$ = 8.2, which is smaller than expected due to the fabrication defect. But note that the reduction of permittivity of dielectric rods does not change the Zak phase and the sign of Im(Z) of each PC, and hence the deterministic edge states still exist. To see this, the projected band structures of two PCs and the edge state dispersions along the $k_y$ direction are shown in Fig. 4(b). Due to the self-complementarity of checkerboard PC, PC1 and PC2 share same projected bulk bands (shaded in black). Inside the common directional band gap (white region), deterministic edge states are found [marked in green in Fig. 4(b)]. As to excite these edge states, a monopole source is placed at the bottom of the boundary (outlined by a green arrow in Fig. 4(a)). It emits electromagnetic waves with various k components and excites the upward edge states along the photonic boundary. As an example, the experimentally measured $E_z$ field of edge state at 7.54 GHz is shown in Fig. 4(e). Here, the measured region is limited to 140×160 mm$^2$ due to the finite size of experimental sample. The electric field is confined at the boundary ($x$ = 70 mm) and decay away from the boundary. It is in good agreement with the numerical simulation result shown in Fig. 4(d). In addition, we perform the fast Fourier transform on the measured fields to retrieve the wave vectors of excited edge states. As shown in Fig. 4(c), the dark and bright colors indicate low and high Fourier amplitudes, respectively. The experimental data (bright color) capture well the numerical dispersions (green line) for the deterministic edge states. Note that edge states around 6.8 GHz are not excited due to the small group velocity for edge states near $k_y$ from 0 to $0.2\pi/a$.

## IV. CONCLUSION

In conclusion, by employing the self-complementary feature of the checkerboard PC, the "a common band gap" condition for finding edge states is naturally satisfied between two PCs which are constructed by different unit cells. Inside the first lowest common band gap, the deterministic edge states found under both the Zak phase analysis and the surface impedance calculation. Lastly, a microwave experiment is also carried out to demonstrate these deterministic edge states.

## ACKNOWLEDGEMENTS

This work is supported by Natural Science Foundation of China (11522437, 11704422, 61775243), Guangdong Natural Science Funds for Distinguished Young Scholar (S2013050015694), Guangdong special support program, and Fundamental Research Funds for the Central Universities (No.17lgpy19).

## REFERENCES

[1] E. Yablonovitch, "Inhibited Spontaneous Emission in Solid-State Physics and Electronics," Physical Review Letters **58**, 2059 (1987).
[2] S. John, "Strong localization of photons in certain disordered dielectric superlattices," Physical Review Letters **58**, 2486 (1987).
[3] J. D. Joannopoulos, S. G. Johnson, J. N. Winn, and R. D. Meade, *Photonic crystals - molding the flow of light* (Princeton University Press, Princeton, NJ, 2008).

[4] H. Kosaka, T. Kawashima, A. Tomita, M. Notomi, T. Tamamura, T. Sato, and S. Kawakami, "Photonic crystals for micro lightwave circuits using wavelength-dependent angular beam steering," Applied Physics Letters **74**, 1370 (1999).

[5] M. Notomi, "Theory of light propagation in strongly modulated photonic crystals: refractionlike behavior in the vicinity of the photonic band gap," Physical Review B **62**, 10696 (2000).

[6] Y. Poo, C. He, C. Xiao, M. H. Lu, R. X. Wu, and Y. F. Chen, "Manipulating one-way space wave and its refraction by time-reversal and parity symmetry breaking," Scientific reports **6**, 29380 (2016).

[7] C. Luo, S. G. Johnson, J. D. Joannopoulos, and J. B. Pendry, "All-angle negative refraction without negative effective index," Physical Review B **65**, 201104 (2002).

[8] Y. Shen, D. Ye, I. Celanovic, S. G. Johnson, J. D. Joannopoulos, and M. Soljačić, "Optical broadband angular selectivity," Science **343**, 1499-1501 (2014).

[9] S.-Y. Lin, E. Chow, V. Hietala, P. R. Villeneuve, and J. D. Joannopoulos, "Experimental demonstration of guiding and bending of electromagnetic waves in a photonic crystal," Science **282**, 274-276 (1988).

[10] I. Söllner, S. Mahmoodian, S. L. Hansen, L. Midolo, A. Javadi, G. Kiršanskė, T. Pregnolato, H. El-Ella, E. H. Lee, J. D. Song, S. Stobbe, and P. Lodahl, "Deterministic photon-emitter coupling in chiral photonic circuits," Nature Nanotechnology **10**, 775 (2015).

[11] M. Xiao, Z. Q. Zhang, and C. T. Chan, "Surface impedance and bulk band geometric phases in one-dimensional systems," Physical Review X **4**, 021017 (2014).

[12] Q. Wang, M. Xiao, H. Liu, S. Zhu, and C. T. Chan, "Measurement of the Zak phase of photonic bands through the interface states of a metasurface/photonic crystal," Physical Review B **93**, 041415 (2016).

[13] X. Shi, C. Xue, H. Jiang, and H. Chen, "Topological description for gaps of one-dimensional symmetric all-dielectric photonic crystals," Optics Express **24**, 18580 (2016).

[14] Y. Yang, T. Xu, Y. F. Xu, and Z. H. Hang, "Zak phase induced multiband waveguide by two-dimensional photonic crystals," Optics Letters **42**, 3085-3088 (2017).

[15] F. Haldane, and S. Raghu, "Possible realization of directional optical waveguides in photonic crystals with broken time-reversal symmetry," Physical Review Letters **100**, 013904 (2008).

[16] Z. Wang, Y. Chong, J. Joannopoulos, and M. Soljačić, "Reflection-free one-way edge modes in a gyromagnetic photonic crystal," Physical Review Letters **100**, 013905 (2008).

[17] Y. Poo, R.-x. Wu, Z. Lin, Y. Yang, and C. T. Chan, "Experimental realization of self-guiding unidirectional electromagnetic edge states," Physical Review Letters **106**, 093903 (2011).

[18] S. A. Skirlo, L. Lu, Y. Igarashi, Q. Yan, J. Joannopoulos, and M. Soljačić, "Experimental observation of large Chern numbers in photonic crystals," Physical Review Letters **115**, 253901 (2015).

[19] A. B. Khanikaev, S. H. Mousavi, W.-K. Tse, M. Kargarian, A. H. MacDonald, and G. Shvets, "Photonic topological insulators," Nature Materials **12**, 233 (2012).

[20] W.-J. Chen, S.-J. Jiang, X.-D. Chen, B. Zhu, L. Zhou, J.-W. Dong, and C. T. Chan, "Experimental realization of photonic topological insulator in a uniaxial metacrystal waveguide," Nature communications **5**, 5782 (2014).

[21] T. Ma, A. B. Khanikaev, S. H. Mousavi, and G. Shvets, "Guiding electromagnetic waves around sharp corners: topologically protected photonic transport in metawaveguides," Physical Review Letters **114**, 127401 (2015).

[22] L.-H. Wu, and X. Hu, "Scheme for achieving a topological photonic crystal by using dielectric

material," Physical Review Letters **114**, 223901 (2015).
[23] T. Ma, and G. Shvets, "All-si valley-Hall photonic topological insulator," New Journal of Physics **18**, 025012 (2016).
[24] X.-D. Chen, F.-L. Zhao, M. Chen, and J.-W. Dong, "Valley-contrasting physics in all-dielectric photonic crystals: Orbital angular momentum and topological propagation," Physical Review B **96**, 020202(R) (2017).
[25] M. J. Collins, F. Zhang, R. Bojko, L. Chrostowski, and M. C. Rechtsman, "Integrated optical Dirac physics via inversion symmetry breaking," Physical Review A **94**, 063827 (2016).
[26] R. K. Pal, and M. Ruzzene, "Edge waves in plates with resonators: an elastic analogue of the quantum valley Hall effect," New Journal of Physics **19**, 025001 (2017).
[27] L. Lu, J. D. Joannopoulos, and M. Soljačić, "Topological photonics," Nature Photonics **8**, 821 (2014).
[28] X.-C. Sun, C. He, X.-P. Liu, M.-H. Lu, S.-N. Zhu, and Y.-F. Chen, "Two-dimensional topological photonic systems," Progress in Quantum Electronics **55**, 52 (2017).
[29] Y. Wu, C. Li, X. Hu, Y. Ao, Y. Zhao, and Q. Gong, "Applications of Topological Photonics in Integrated Photonic Devices," Advanced Optical Materials, 1700357 (2017).
[30] X. Huang, Y. Yang, Z. H. Hang, Z.-Q. Zhang, and C. T. Chan, "Geometric phase induced interface states in mutually inverted two-dimensional photonic crystals," Physical Review B **93**, 085415 (2016).
[31] X. Huang, M. Xiao, Z.-Q. Zhang, and C. T. Chan, "Sufficient condition for the existence of interface states in some two-dimensional photonic crystals," Physical Review B **90**, 075423 (2014).
[32] Y. Yang, X. Huang, and Z. H. Hang, "Experimental Characterization of the Deterministic Interface States in Two-Dimensional Photonic Crystals," Physical Review Applied **5**, 034009 (2016).
[33] M. Senechal, "Color symmetry," Computer & Mathematics with Applications **16**, 545 (1988).
[34] W. Kohn, "Analytic properties of Bloch waves and Wannier Functions," Physical Review A **115**, 809 (1959).
[35] J. W. Dong, J. Zeng, Q. F. Dai, and H. Z. Wang, "Universal condition for the existence of interface modes in the whole momentum space with arbitrary materials," Arxiv **0801.4117** (2008).
[36] X. Chen, T. M. Grzegorczyk, B.-I. Wu, J. Pacheco, and J. A. Kong, "Robust method to retrieve the constitutive effective parameters of metamaterials," Physical Review E **70**, 016608 (2004).
[37] D. R. Smith, D. C. Vier, T. Koschny, and C. M. Soukoulis, "Electromagnetic parameter retrieval from inhomogeneous metamaterials," Physical Review E **71**, 036617 (2005).
[38] Y. Zhou, X.-T. He, F.-L. Zhao, and J.-W. Dong, "Proposal for achieving in-plane magnetic mirrors by silicon photonic crystals," Optics Letters **41**, 2209 (2016).

# FIGURES AND FIGURE CAPTIONS

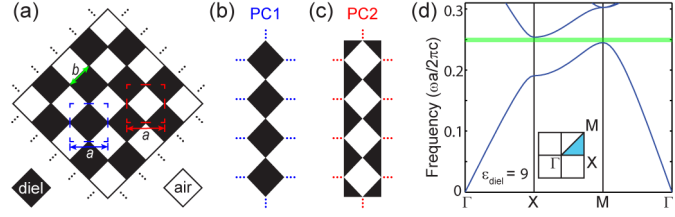

FIG. 1. (Color online) (a) The schematic of checkerboard PC which consists of two interlaced square lattices of dielectric rods with $\varepsilon_{diel}$ (black patches) and air rods (white patches). The lattice constant of PC is $a$, and the side length of dielectric and air rods is $b = a/\sqrt{2}$. Due to the self-complementarity of checkerboard PC, the unit cell can be chosen to be centered at dielectric rod (blue dashed square) or air rod (red dashed square). (b, c) The schematics of (b) PC1 which is constructed by periodically repeating the dielectric rod centered unit cell and (c) PC2 which is constructed by periodically repeating the air rod centered unit cell. (d) Band structure for the transverse magnetic modes of the checkerboard PC with $\varepsilon_{diel} = 9$. The inset shows the first Brillouin zone. A complete band gap ranging from $0.245c/a$ to $0.254c/a$ is highlighted by a green rectangle. This band gap is commonly shared by PC1 and PC2 due to the self-complementarity of checkerboard PC.

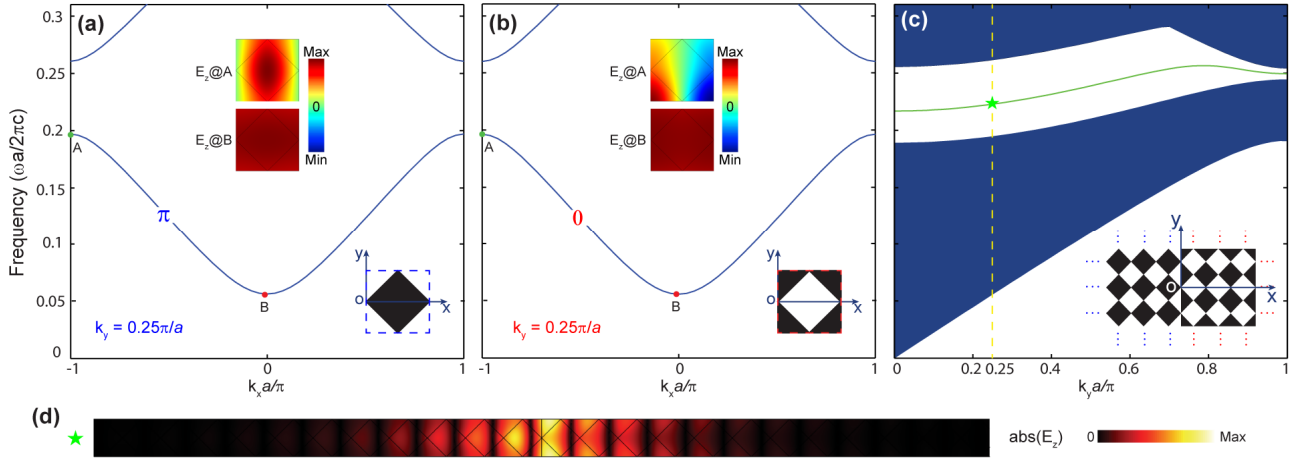

FIG. 2. (Color online) Zak phase analysis. (a) Band structure, eigen-fields and Zak phase for PC1 whose unit cell is centered at dielectric rod. The schematic of the unit cell is shown in the right inset. The coordinate origin is located at the center of the left boundary of the unit cell. The middle inset shows the electric fields of two eigen-states at high symmetry A ($k_x = -\pi/a$) and B ($k_x = 0$) points. As $|\tilde{E}_A(x=0)| = 0$ and $|\tilde{E}_B(x=0)| \neq 0$, the Zak phase is $\pi$ which is given along with the reduced 1D band structure for $k_y = 0.25\pi/a$. (b) Band structure, eigen-fields and Zak phase for PC2 whose unit cell is centered at air rod. Due to different field distributions of eigen-states at A or B points comparing to those of PC1, the Zak phase for PC2 changes to be 0. (c) Projected band structures for the photonic boundary between semi-infinite PC1 and PC2. The green line represents edge states dispersion, and the low right inset shows the schematic of photonic boundary. (d) The amplitude of electric fields of one representative edge state with $f = 0.223c/a$ at $k_y = 0.25\pi/a$ (marked by a green star in (c)). Fields focus near the boundary.

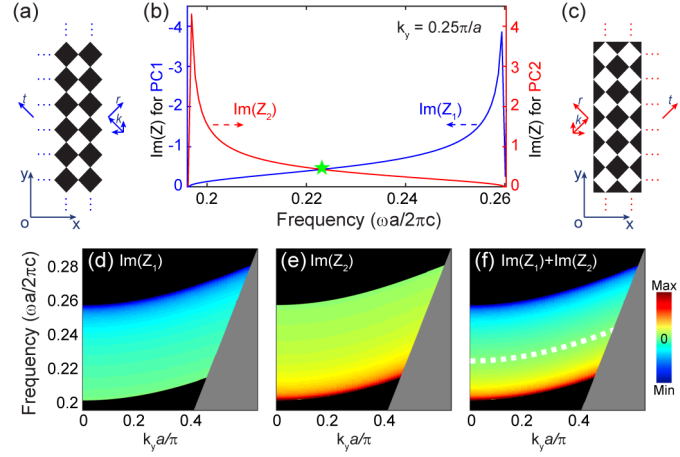

FIG. 3. (Color online) Surface impedance calculation. The schematics of (a) the semi-infinite PC1 and (c) the semi-infinite PC2 for the transmission simulation. (b) The value of Im(Z) for PC1 (i.e., Im($Z_1$) and marked in blue) and PC2 (i.e., Im($Z_2$) and marked in red) as a function of frequency. The tangential component of incident wave-vector is $k_y = 0.25\pi/a$. One edge state exists at $f = 0.223c/a$ at which the zero impedance condition Im($Z_1$) + Im($Z_2$) = 0 is fulfilled (green star). (d) Distribution of surface impedance Im($Z_1(\omega, k_y)$) for PC1. (e) Distribution of surface impedance Im($Z_2(\omega, k_y)$) for PC2. (f) Distribution of Im($Z_1$) + Im($Z_2$). For each $k_y$, Im($Z_1$) + Im($Z_2$) drops monotonically from $+\infty$ to $-\infty$ with the increasing frequency, and goes across the value of zero. A white-dashed line outlines the positions ($\omega, k_y$) where Im($Z_1$) + Im($Z_2$) = 0 is fulfilled.

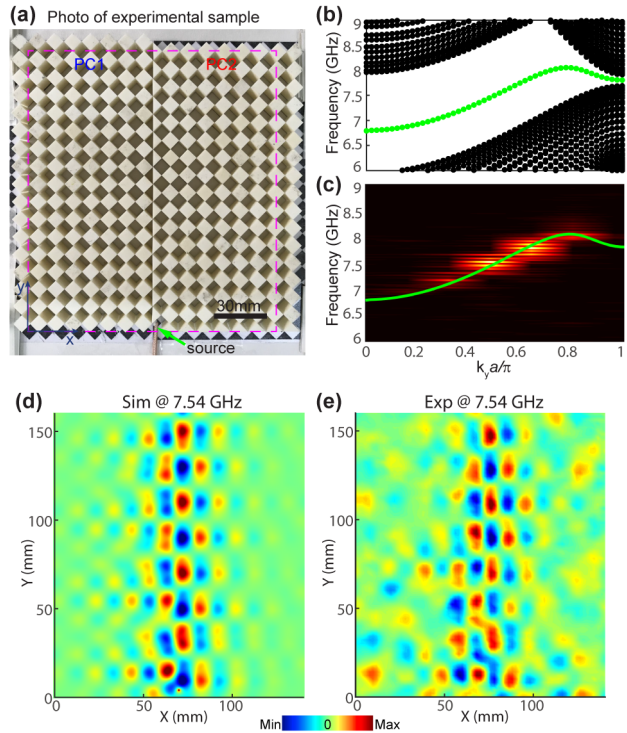

FIG. 4. (Color online) Microwave experimental verification. (a) Photo of experimental sample with PC1 on the left while PC2 on the right. The dielectric rod is alumina rod with $\varepsilon_{\text{diel}} = 8.2$, and the lattice constant of PC is $a = 10$ mm. A pink dashed rectangle outlines the scanned region in the experiment, and a green arrow marks the monopole source. (b) Calculated projected bands (black) and edge states (green) dispersions for the boundary shown in (a). (c) Measured dispersions (bright color) for the deterministic edge states, compared with the numerical data (green line). (d) The simulated and (e) measured $E_z$ fields of edge states at the frequency of 7.54 GHz.